\documentclass[a4paper,preprint,preprintnumbers,nofootinbib]{revtex4}
\usepackage[dvipdf,dvips,dvipdfmx]{graphicx}
\usepackage{amsmath}
\usepackage{amssymb}
\usepackage{float}
\usepackage{wrapfig}
\usepackage{bm}

\begin{document}

\title{Tensor renormalization group analysis of ${\rm CP}(N-1)$ model}

\author{Hikaru Kawauchi}
\email{kawauchi@hep.s.kanazawa-u.ac.jp}
\author{Shinji Takeda}
\email{takeda@hep.s.kanazawa-u.ac.jp}
\affiliation{Institute for Theoretical Physics, Kanazawa University, Kanazawa 920-1192, Japan}

\date{\today}

\preprint{KANAZAWA-16-02}

\begin{abstract}
We apply the higher-order tensor renormalization group to the lattice CP($N-1$) model in two dimensions. 
A tensor network representation of the CP($N-1$) model in the presence of the $\theta$ term is derived. 
We confirm that the numerical results of the CP(1) model without the $\theta$ term using this method are consistent with that of the O(3) model which is analyzed by the same method in the region $\beta \gg 1$ and that obtained by the Monte Carlo simulation in a wider range of $\beta$. The numerical computation including the $\theta$ term is left for future challenges.
\end{abstract}
\maketitle

\section{Introduction}

One possible way to avoid the sign problem is to simply abandon Monte Carlo simulation. Levin and Nave proposed the tensor renormalization group (TRG) in Ref.~\cite{Michael}, where they study two-dimensional classical systems. This method has no sign problem because it does not consider the Boltzmann weight as a probability of generating field configurations. This method makes it possible to calculate the partition function approximately based on the singular value decomposition. The TRG method comprises mainly two steps. The first step is to construct a tensor network representation of the partition function. The detail is described in the next section. The second step is to decrease the number of tensors under control of systematic errors. After that, one can finally compute the partition function approximately by contracting all indices of a few coarse grained tensors. While the original TRG was a method studying two-dimensional systems, Xie {\it et al.}\cite{Xie} introduced the higher-order TRG (HOTRG) as an extension to higher-dimensional systems, which is based on the higher-order singular value decomposition.

One example of systems having the sign problem is a system including the $\theta$ term. The QCD Lagrangian naturally includes this term, which breaks CP symmetry. $\theta$ is a free parameter; for example, it takes from 0 to 2$\pi$. From the experiment of the neutron electric dipole moment, however, the parameter has the upper bound $|\theta| < 10^{-10}$ \cite{Baker, Crewther}.  Why is the free parameter $\theta$ so small? This puzzle is the so-called strong CP problem. In order to answer the question, lattice QCD simulation including the $\theta$ term is desirable, but the presence of the $\theta$ term causes the sign problem. 

Instead of approaching QCD directly, it is reasonable to start to investigate its toy model, the CP($N-1$) model, which shares many features with QCD and has the $\theta$ term too. A long time ago, Schierholz suggested an interesting scenario to solve the strong CP problem in the CP($N-1$) model by analyzing the phase diagram in the $\beta$ - $\theta$ plane \cite{Schierholz}. In the phase diagram, there are two phases, the confining phase and deconfinement phase (Higgs phase), which are distinguished by observing ``flattening'' of free energy, and it was argued that in the continuum limit, if $\theta=0$ is the only choice in the confining phase, the strong CP problem would be resolved.
To confirm this result, the authors of Refs.~\cite{PlefkaMonte, Imachi} also investigated the similar phase diagram by a similar way as Schierholz, where they used the Monte Carlo method and computed the partition function in the presence of the $\theta$ term $Z(\theta)$ by computing the topological charge distribution $P(Q)$. However, they insisted that the main reason of the ``flattening'' of free energy is due to the statistical fluctuation of $P(Q)$ in their analysis.
On the other hand, Plefka and Samuel \cite{Plefka} supported Schierholz's result by using another method, the strong coupling expansion, but they could not confirm whether $\theta$ must be taken to zero in the continuum limit since they truncated higher-order corrections.
Although it is not clear that the solution can be directly applied to QCD,
it is interesting to verify the scenario with another method, namely the TRG approach which is free of the sign problem. Moreover, this method can drastically reduce the truncation error in an expansion of $\beta$.

In particular, the CP(1) model, which is equivalent to the O(3) model, is interesting since it is connected with a verification of the Haldane conjecture \cite{Haldane:1982rj,Haldane:1983ru}: the two-dimensional O(3) nonlinear sigma model with $\theta=\pi$ is gapless.
In order to prove this conjecture, Monte Carlo simulations for the O(3) model with the $\theta$ term were conducted in various methods.
In spite of the presence of the sign problem, the extended cluster algorithm \cite{Bietenholz:1995zk,Bogli:2011aa,deForcrand:2012se} and the refined analysis \cite{Alles:2007br,Azcoiti:2012ws,Alles:2014tta} made the calculation feasible.
As a result, the second-order phase transition at $\theta=\pi$ was found
and it was shown that the observed critical exponents belong to the universality class of the Wess-Zumino-Novikov-Witten (WZNW) model with a topological coupling $k=1$ \cite{Wess:1971yu,Novikov:1981,Witten:1983ar} as expected \cite{Affleck:1987ch,Affleck:1988px}.
Furthermore, the phase structure of the two-dimensional CP(1) model in the presence of the $\theta$ term was also investigated by the Monte Carlo simulation \cite{Azcoiti:2007cg}.
The universality class, however, turned out to be different from that of the WZNW model.
Therefore it is worthwhile to assess the universality class and reanalyze the phase structure by using the sign-problem-free method, i.e., tensor renormalization group methods.

Our purpose in this paper is to apply the HOTRG to the CP($N-1$) model in two dimensions. The organization of this paper is as follows: In Sec.~\ref{2}, we present a tensor network representation of the CP($N-1$) model including the $\theta$ term, in Sec.~\ref{3} we show the numerical results at $\theta=0$, and in Sec.~\ref{4} we give our conclusions.

\section{Tensor network representation of the CP($N-1$) model}\label{2}
The partition function of the lattice CP($N-1$) model in two dimensions as a function of inverse coupling constant $\beta$ and the parameter $\theta$ is given by
\begin{align}
Z=
\int 
\prod_{x} dz(x) 
\prod_{x, \mu}dU_{\mu}(x)~
{\rm e}^{-S_\theta},
\end{align}
where
\begin{align}
S_\theta=
-\beta N\sum_{x, \mu}
\Bigl[
z^\ast(x) \cdot z(x+\hat{\mu})U_{\mu}(x)+
z(x) \cdot z^\ast(x+\hat{\mu}) U_{\mu}^\dag(x)
\Bigr]
-
i\frac{\theta}{2\pi}\sum_{p} q_p,
\end{align}
\begin{align}
dz(x)\equiv
d^Nz(x) d^Nz^\ast(x) \delta(|z(x)|-1),
\end{align}
and $z^a(x)$ is the $N$-component complex scalar field ($a=1,\cdots, N$) of unit length, $z(x)\cdot z(x)\equiv z^{\ast a}(x)z^a(x)=1$, and $U_{\mu}(x)$ is the link variable described by the auxiliary vector field $A_{\mu}(x)$, i.e., $U_{\mu}(x)={\rm exp}\{iA_{\mu}(x)\}$. The second term in the action is the $\theta$ term and 
\begin{align}
q_p=A_{1}(x)+A_{2}(x+\hat{1})-A_{1}(x+\hat{2})-A_{2}(x)~~{\rm mod}~2\pi.
\end{align}
In order to obtain a tensor network representation, one has to expand the Boltzmann weight with new integers and then integrate out the old degrees of freedom (the complex fields $z(x)$ and the auxiliary field $A_\mu(x)$ in this case). In the end, one can obtain a tensor which has indices of the new integers.

First, we derive a tensor network representation without the $\theta$ term and then consider the $\theta$ term. To expand the Boltzmann weight with new integers, we use the characterlike expansion \cite{Plefka},
\begin{align}
\nonumber
&{\rm exp}\left\{
\beta N
\Bigl[
z^\ast (x)\cdot z(x+\hat{\mu}) U_{\mu}(x)+
z(x) \cdot z^\ast(x+\hat{\mu}) U_{\mu}^\dag(x)
\Bigr]
\right\}\\
&=
Z_0(\beta)\sum_{l,m=0}^\infty d_{(l;m)}{\rm exp}[i(m-l)A_{\mu}(x)]h_{(l;m)}(\beta)
f_{(l;m)}(z(x), z(x+\hat{\mu})), 
\label{charactercpn}
\end{align}
where $d_{(l;m)}$ are dimensionalities of characterlike representations, $h_{(l;m)}(\beta)$ are characterlike expansion coefficients, $f_{(l;m)}(z(x), z(x+\hat{\mu}))$ are characterlike expansion characters, and $Z_0(\beta)$ is the normalization factor which makes $h_{(0;0)}(\beta)=1$. The integers $l$ and $m$ are non-negative and will become the indices of the tensor shown below.

The characterlike expansion coefficients $h_{(l;m)}(\beta)$ are expressed by the modified Bessel functions of the first kind,
\footnote{We confirm this expression for any integer $l \geq 0$ and $m=0, 1, 2, 3$.}
\begin{align}
h_{(l;m)}(\beta)=\frac{I_{N-1+l+m}(2N\beta)}{I_{N-1}(2N\beta)}.
\label{hbeta}
\end{align}
Since the modified Bessel function of the first kind, $I_n(x)$, decreases rapidly as $n$ increases with a fixed value of $x$, one can safely truncate the sum of $l$ and $m$ in Eq. (\ref{charactercpn}) at some order (say $l=m=l_{\rm max}$).
The normalization factor is given by
\begin{align}
Z_0(\beta)=
\frac{(N-1)!I_{N-1}(2N\beta)}{(N\beta)^{N-1}}.
\label{Z0}
\end{align}

In Appendix \ref{characters}, we show some explicit form
\footnote{From these examples, $f_{(l;m)}(z(x),z(x+\hat{\mu}))$ for any $l$ and $m$ is inferred.}
 of the dimensionalities of characterlike representations $d_{(l;m)}$ and the characterlike expansion characters $f_{(l;m)}(z(x),z(x+\hat{\mu}))$ for any integer $l \geq 0$. The $d_{(l;m)}$ and $f_{(l;m)}$ are determined so as to satisfy the following conditions,
\begin{align}
\int dz~
f_{(l;m)}(z^\prime, z)
f^\ast_{(l';m')}(z^{\prime\prime}, z)
=
\frac{1}{d_{(l;m)}}\delta_{l,l'}\delta_{m,m'}
f_{(l;m)}(z^\prime, z^{\prime\prime}),
\end{align}
\begin{align}
f_{(l;m)}(z, z)
=d_{(l;m)}.
\end{align}

\begin{figure}[t]
 %\hspace{22mm}
 \centering
  \includegraphics[width=65mm]{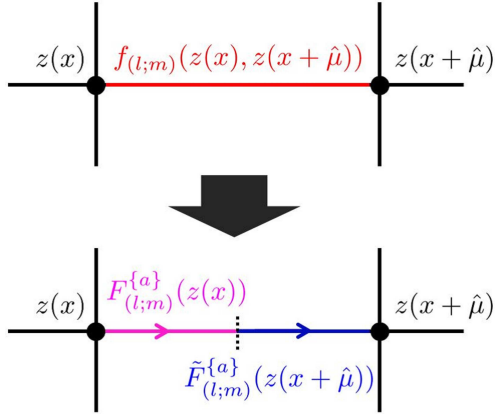}
  %\hspace{-10mm}
  %\vspace{15mm}
 \caption{Decomposition of characterlike expansion characters $f_{(l;m)}(z(x),z(x+\hat{\mu}))$.}
 \label{cpndecompose}
\end{figure}

The term, $f_{(l;m)}(z(x),z(x+\hat{\mu}))$, is expressed by the combination of two complex scalar fields, $z(x)$ and $z(x+\hat{\mu})$. In order to obtain a tensor network representation, one has to integrate out the complex scalar fields $z$ site by site.
For that purpose, it is convenient to rewrite it as follows,
\begin{align}
f_{(l;m)}(z(x),z(x+\hat{\mu}))
=
\sum_{\{a\}}
F^{\{a\}}_{(l;m)}(z(x))
\tilde{F}^{\{a\}}_{(l;m)}(z(x+\hat{\mu})),
\end{align}
where $\{a\}=a_1, a_2, \cdots, a_l, a'_1, a'_2, \cdots, a'_m$, and
$a_i=1,2,\cdots, N$ for
$i=1, 2, \cdots, l$, 
$a'_j=1,2,\cdots, N$ for
$j=1, 2, \cdots, m$.
A pictorial expression of this decomposition is given in Fig.~\ref{cpndecompose}.
The explicit forms of $F$ and $\tilde F$ are shown in Appendix \ref{FFt}.

\begin{figure}[h]
 %\hspace{22mm}
 \centering
  \includegraphics[width=120mm]{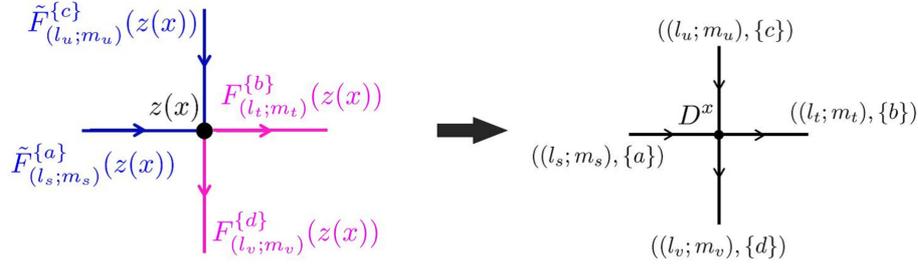}
  %\hspace{-10mm}
  %\vspace{15mm}
 \caption{Integrating out the $N$-component complex scalar field $z(x)$.}
 \label{integrateout}
\end{figure}

After the decomposition, the last step is to integrate out the old degrees of freedom, $z$ and $A$.
If we focus on a site $x$, there are two $F$s and two $\tilde{F}$s, as illustrated in Fig.~\ref{integrateout}. 
%A tensor network representation is given by the integration over $z_i$ and $z_i^\ast$, 
A tensor expressed in terms of them is given by
\begin{align}
\nonumber
&D^x_{((l_s;m_s),\{a\})((l_t;m_t),\{b\})((l_u;m_u),\{c\})((l_v;m_v),\{d\})}\\
\nonumber
=&\int dz(x) 
\sqrt{d_{(l_s;m_s)}d_{(l_t;m_t)}d_{(l_u;m_u)}d_{(l_v;m_v)}
h_{(l_s;m_s)}(\beta)h_{(l_t;m_t)}(\beta)
h_{(l_u;m_u)}(\beta)h_{(l_v;m_v)}(\beta)}\\
&\hspace{50pt}
\times
\tilde{F}^{\{a\}}_{(l_s;m_s)}(z(x))
F^{\{b\}}_{(l_t;m_t)}(z(x))
\tilde{F}^{\{c\}}_{(l_u;m_u)}(z(x))
F^{\{d\}}_{(l_v;m_v)}(z(x)).
\label{dtensor}
\end{align}

\noindent
The integration of the complex scalar fields, $z(x)$ and $z^\ast (x)$, can be done analytically thanks to the integral
\cite{Samuel,Plefka}

\begin{align}
\nonumber
\int dz~ z^{a_1}z^{a_2}\cdots z^{a_m}z^{\ast b_1}z^{\ast b_2}\cdots z^{\ast b_k}
\equiv&
\int d^Nz d^Nz^\ast \delta(|z|-1)
 z^{a_1}z^{a_2}\cdots z^{a_m}z^{\ast b_1}z^{\ast b_2}\cdots z^{\ast b_k}\\
\nonumber
=&
({\rm const.})\times
\delta^m_k
\frac{(N-1)!}{(N-1+m)!}\sum_{\sigma \in S_m} 
\delta^{a_1}_{b_{\sigma_1}} \delta^{a_2}_{b_{\sigma_2}}\cdots  \delta^{a_m}_{b_{\sigma_m}}\\
\equiv&
({\rm const.})\times
\delta^m_k
\frac{(N-1)!}{(N-1+m)!}
\delta^{\{a_1, a_2, \cdots , a_{m}\}}
_{\{b_1, b_2, \cdots , b_{m}\}},
\label{zint}
\end{align}
where the sum means all permutations of the indices $b$. The detail of the tensor $D$ is shown in Appendix \ref{Dtensor}.

\begin{figure}[h]
 \centering
  \includegraphics[width=40mm]{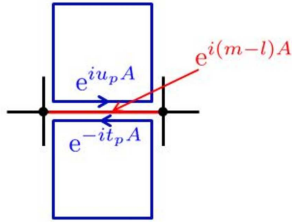}
 \caption{Integrating out the link variable.}
 \label{Aintegration}
\end{figure}

At this point, we mention the integration of the link variable. The character expansion of the $\theta$ term \cite{Hassan:1995dn,Shimizu:2014fsa} is given by

\begin{align}
\nonumber
{\rm e}^{i\frac{\theta}{2\pi}q_p}
&=
\sum_{n_p\in \mathbb{Z}}
{\rm e}^{in_p \varphi_p}\frac{2{\rm sin}
\frac{\theta+2\pi n_p}{2}}{\theta+2\pi n_p}
\hspace{40pt} (\varphi_p \equiv A_{1}(x)+A_{2}(x+\hat{1}) - A_{1}(x+\hat{2})-A_{2}(x))
\\
\nonumber
&\equiv
\sum_{n_p\in \mathbb{Z}}
{\rm e}^{in_p \varphi_p}
G_{n_p}(\theta)\\
&=
\sum_{n_p\in \mathbb{Z}}
{\rm e}^{in_p (A_{1}(x)+A_{2}(x+\hat{1}) - A_{1}(x+\hat{2})-A_{2}(x))
}
G_{n_p}(\theta).
\end{align}
In contrast to the case of the complex scalar fields, the integration of the link variable is rather simple. For one link variable, four integers are coupled (Fig.~\ref{Aintegration}):
\begin{align}
\nonumber
\int_{-\pi}^{\pi} dA \ 
{\rm e}^{i(u_p-t_p)A}
{\rm e}^{i(m-l)A}
&=
\delta^{l-m}_{u_p-t_p}\\
&\equiv
H^{(l;m)}_{u_p,t_p}.
\end{align}

\noindent
If $\theta=0$, the indices are selected to $u_p=t_p=0$ and the tensor $H$ just gives a constraint that the integer $l$ is equivalent to the integer $m$, i.e.,
\begin{align}
G_{n_p}(\theta =0) =
\frac{{\rm sin}{\pi n_p}}{\pi n_p}
= \delta_{n_p, 0},
\end{align}
\begin{align}
H^{(l;m)}_{u_p,t_p}\to H^{(l;m)}_{0,0}=\delta_{l,m}.
\end{align}

\begin{figure}[h]
 \centering
  \includegraphics[width=120mm]{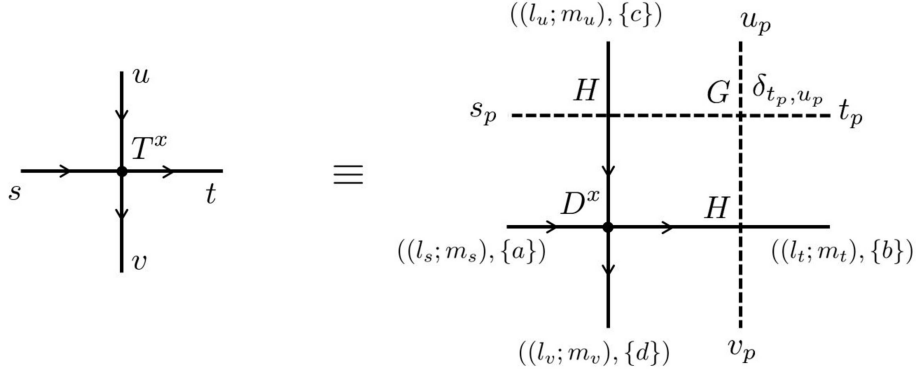}
 \caption{Tensor of the CP($N-1$) model.}
 \label{cpntensor}
\end{figure}

The combination of these tensors derived above leads to one tensor network representation,

\begin{align}
\nonumber
T^x_{stuv}
\equiv&T^x_{((l_s;m_s),\{a\},s_p)((l_t;m_t),\{b\},t_p)
((l_u;m_u),\{c\},u_p)((l_v;m_v),\{d\},v_p)}\\
\nonumber
\equiv&D^x_{((l_s;m_s),\{a\})((l_t;m_t),\{b\})
((l_u;m_u),\{c\})((l_v;m_v),\{d\})}\\
&\times H^{(l_t;m_t)}_{u_p,v_p}H^{(l_u;m_u)}_{s_p,t_p}
G_{t_p}(\theta)\delta_{t_p,u_p}.
\label{tensor}
\end{align}
This combination of tensors is depicted in Fig.~\ref{cpntensor}.

%\footnote{}

\section{Numerical results}\label{3}
\begin{figure}[h]
 \centering
  \includegraphics[width=12cm]{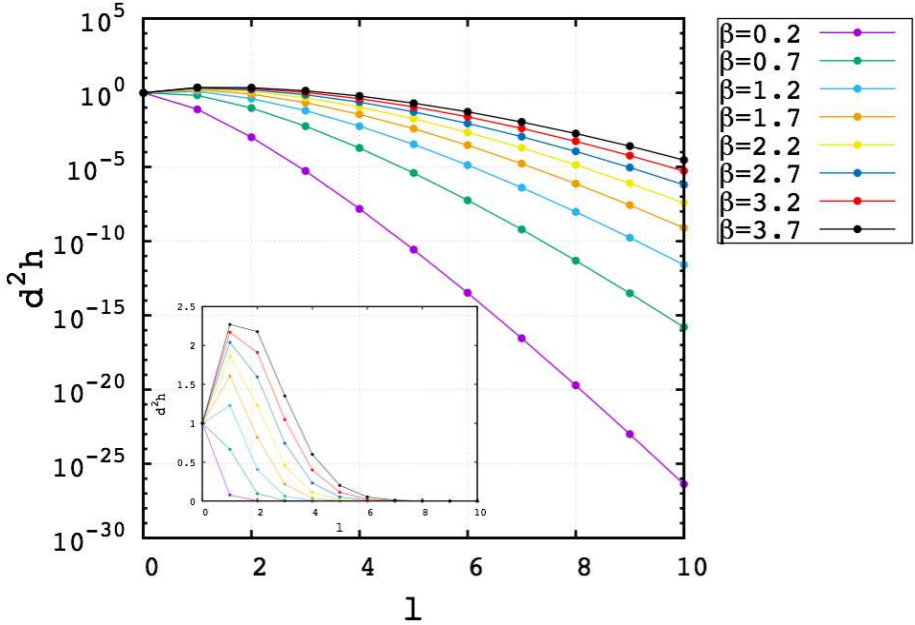}
 \caption{The weights $d_{(l;m)}^2h_{(l;m)}(\beta)$ of the characterlike expansion [Eq. (\ref{charactercpn})] in the case $N=2$ and $l=m$. We suppose that $d_{(l;l)}=\sqrt{2l+1}$ and $h_{(l;l)}(\beta)$ in Eq. (\ref{hbeta}) for any $l$.}
 \label{weight}
\end{figure}

By using the tensor in Eq. (\ref{tensor}), we apply the HOTRG and obtain the partition function of the CP($N-1$) model in the case $N=2$ and $\theta=0$, in which case the tensor is given by
\begin{align}
\nonumber
T^x_{stuv}
\equiv&T^x_{((l_s;l_s),\{a\},0)((l_t;l_t),\{b\},0)
((l_u;l_u),\{c\},0)((l_v;l_v),\{d\},0)}\\
=&D^x_{((l_s;l_s),\{a\})((l_t;l_t),\{b\})
((l_u;l_u),\{c\})((l_v;l_v),\{d\})},
\end{align}
where each $a_i$ and $a'_j$ take $1$ or $2$ for $i=1, 2, \cdots, l_s$ and $j=1, 2, \cdots, l_s$ and the same holds for $\{b\}$, $\{c\}$ and $\{d\}$. Since this tensor has infinite number of elements, we calculate for two cases: (i) $l=0,1$ ($l_{\rm max}=1$) and (ii) $l=0,1,2$ ($l_{\rm max}=2$), which correspond to the truncation of the sum of the integers $l$ and $m$ in Eq. (\ref{charactercpn}). Due to the truncation, the bond dimension of the tensor is $(4^{l_{\rm max}+1}-1)/3$.
One can see the weights $d_{(l;l)}^2h_{(l;l)}(\beta)$ in Fig.~\ref{weight}.
\footnote{Here, we put the value of $f_{(l;m)}(z(x),z(x+\hat{\mu}))$ as its maximum, $d_{(l;m)}$.} 
For $N=2$, $d_{(l;l)}$ equals $\sqrt{2l+1}$.
\footnote{We confirm this expression for the case $l =0, 1, 2, 3$.}
This figure indicates that the truncation error grows as $\beta$ increases.

\begin{figure}[h]
 %\hspace{22mm}
 \centering
  \includegraphics[width=12cm]{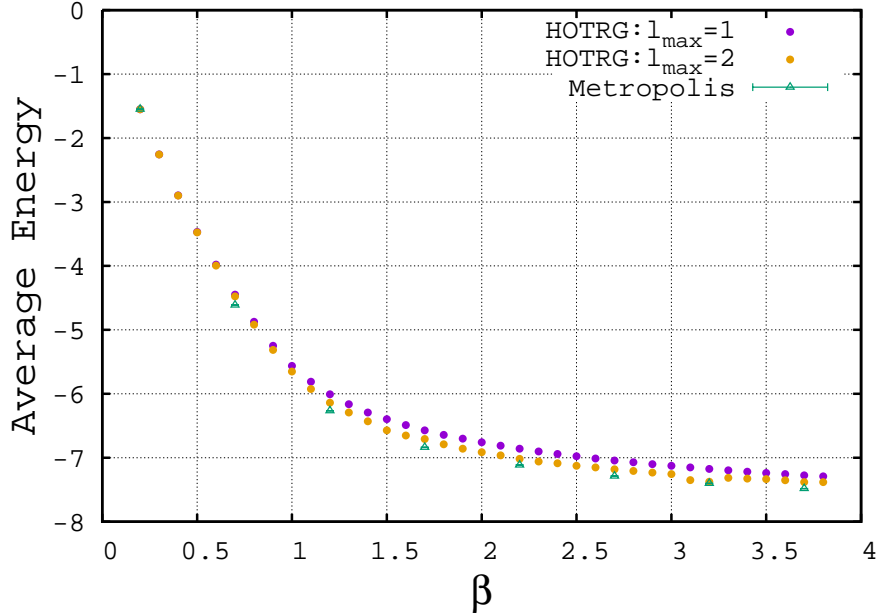}
  %\hspace{-10mm}
  %\vspace{15mm}
 \caption{Average energy of the CP(1) model computed by HOTRG and Metropolis algorithm. The lattice size is $4 \times 4$. The circle marks indicate the results of HOTRG and the triangle marks indicate the results of Metropolis algorithm. }
 \label{hotrgvsmetro}
\end{figure}

First, we compare the result of the TRG method ($l_{\rm max}=1, 2$) with that of the Monte Carlo simulation. Figure~\ref{hotrgvsmetro} compares the average energy of the CP(1) model computed by the two methods on $4\times4$ lattice. The average energy $E$ is defined by
\begin{align}
E=-\frac{1}{L^2}\frac{\partial}{\partial \beta}{\rm ln}Z,
\end{align}
where $L$ is the linear lattice size.
We take the derivative with respect to $\beta$ numerically in the TRG method.
In our Monte Carlo simulation, we use the Metropolis algorithm and $10^6$ configurations after thermalization are generated for each $\beta$. The statistical errors are estimated by the jackknife method and the autocorrelation time $\tau_{\rm int} = 1-65$.
The result of the TRG method ($l_{\rm max}=2$) is almost consistent with that of the Monte Carlo simulation. The little difference between the two results is considered to the truncation error $l_{\rm max}=2$ of the HOTRG. 
It is expected that these two results are consistent at sufficiently large $l_{\rm max}$.

\begin{figure}[h]
 %\hspace{22mm}
 \centering
  \includegraphics[width=12cm]{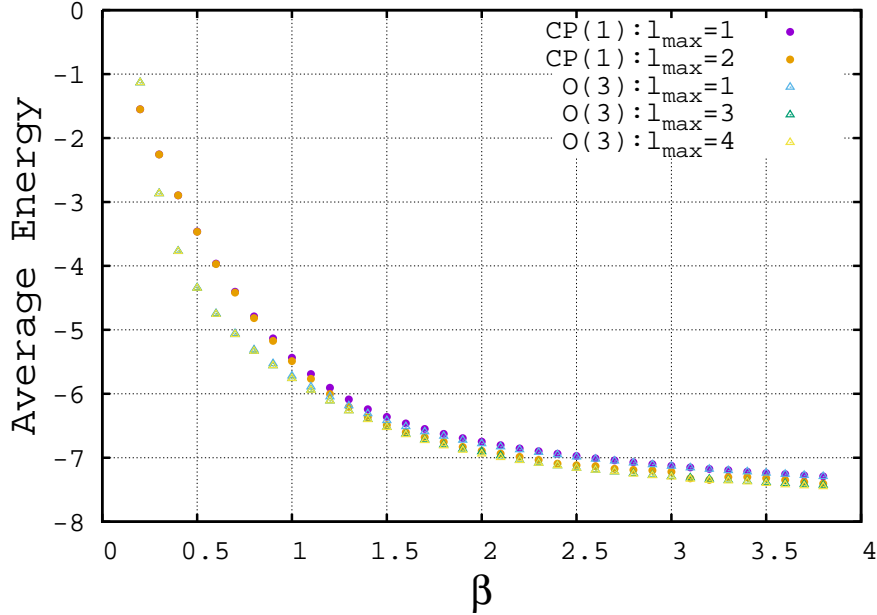}
 %\hspace{-10mm}
  %\vspace{15mm}
 \caption{Average energy of the CP(1) model and the O(3) model computed by using HOTRG. The lattice size is $2^{20} \times 2^{20}$. The circle marks indicate the results of the CP(1) model and the triangle marks indicate the results of the O(3) model.}
 \label{cp1o3}
\end{figure}

Next, Fig.~\ref{cp1o3} compares the result of the HOTRG with that of the O(3) model on $2^{20}\times2^{20}$ lattice which is analyzed by the same method. Unmuth-Yockey {\it et al.} applied the HOTRG to the O(3) model in Ref.~\cite{Unmuth}. By following them, we compute the average energy of the O(3) model. The energy of the two models is connected to each other in the continuum limit by the relation

\begin{align}
\frac{1}{\beta}+E_{\rm O(3)}(\beta)
=
E_{\rm CP(1)}(\beta)+6.
\end{align}
Using this relation, we mapped the result of the O(3) model into the graph. In the limit $\beta \to \infty$, that is, the continuum limit, these two results are expected to be consistent
\footnote{Here, we assume that the truncation error is small in the large $\beta$ region.}
 and in fact such a tendency is observed.

\section{Summary and Outlook}\label{4}
In this work, we show a tensor network representation of the CP($N-1$) model including the $\theta$ term.
It is confirmed that the numerical results of the CP(1) model at $\theta=0$ using the TRG method are consistent with that computed by the Monte Carlo simulation and that of the O(3) model which is analyzed by the same method in the region $\beta \gg 1$.

For our future work, we shall try to do the implementation including the $\theta$ term. In the presence of this term, $l$ no longer equals $m$. In this case, the computational cost of the TRG methods turns out to be very expensive, and we may need some techniques to reduce the cost.

\subsection*{Acknowledgements}
We would like to thank Yoshinobu Kuramashi, Yuya Shimizu, Yusuke Yoshimura, Norihiro Nakamoto, and Ryo Sakai for useful discussion and comments. H. K. is grateful to Yoshifumi Nakamura for advice about computer skills and Yannick Meurice for his helpful advice. This work is supported by Kanazawa University SAKIGAKE Project.

\begin{appendix}

\section{Some examples of dimensionalities of characterlike representations $d$ and characterlike expansion characters $f$}\label{characters}

\noindent
For $m=0$,

\begin{align}
d_{(l;0)}=\sqrt{\frac{(N-1+l)!}{l!(N-2)!(N-1)}},
\end{align}
\begin{align}
f_{(l;0)}(z(x),z(x+\hat{\mu}))=\sqrt{\frac{(N-1+l)!}{l!(N-2)!(N-1)}}(z(x) \cdot z^\ast(x+\hat{\mu}))^l.
\label{fl0}
\end{align}

\noindent
For $m=1$,
\begin{align}
d_{(l;1)}
=\sqrt{\frac{(N+l)!(N-1+l)}{l!(N-1)!(N-1)}}\frac{N-1}{N-1+l},
\end{align}
\begin{align}
\nonumber
&f_{(l;1)}(z(x),z(x+\hat{\mu}))=\\
&\sqrt{\frac{(N+l)!(N-1+l)}{l!(N-1)!(N-1)}}
\Bigl[(z(x)\cdot z^\ast(x+\hat{\mu}))^l(z^\ast(x) \cdot z(x+\hat{\mu}))
-\frac{l}{N-1+l}(z(x) \cdot z^\ast(x+\hat{\mu}))^{l-1}\Bigr].
\label{fl1}
\end{align}

\noindent
For $m=2$,
\begin{align}
d_{(l;2)}=\sqrt{\frac{(N+1+l)!(N+l)(N-1+l)}{2!l!N!(N-1)}}\frac{N(N-1)}{(N+l)(N-1+l)},
\end{align}
\begin{align}
\nonumber
&f_{(l;2)}(z(x),z(x+\hat{\mu}))\\
\nonumber
&=\sqrt{\frac{(N+1+l)!(N+l)(N-1+l)}{2!l!N!(N-1)}}
\Bigl[(z(x) \cdot z^\ast(x+\hat{\mu}))^l(z^\ast(x) \cdot z(x+\hat{\mu}))^2\\
&\hspace{11pt}
-\frac{2l}{N+l}(z(x)\cdot z^\ast(x+\hat{\mu}))^{l-1}(z^\ast(x) \cdot z(x+\hat{\mu}))
+\frac{l(l-1)}{(N+l)(N-1+l)}(z(x)\cdot z^\ast(x+\hat{\mu}))^{l-2}\Bigr].
\label{fl2}
\end{align}

%\begin{comment}
\noindent
For $m=3$,
\begin{align}
d_{(l;3)}=\sqrt{\frac{(N+2+l)!(N+1+l)(N+l)(N-1+l)}{3!l!(N+1)!(N-1)}}\frac{(N+1)N(N-1)}{(N+1+l)(N+l)(N-1+l)},
\end{align}
\begin{align}
\nonumber
&f_{(l;3)}(z(x),z(x+\hat{\mu}))\\
\nonumber
&=\sqrt{\frac{(N+2+l)!(N+1+l)(N+l)(N-1+l)}{3!l!(N+1)!(N-1)}}
\Bigl[(z(x) \cdot z^\ast(x+\hat{\mu}))^l(z^\ast(x) \cdot z(x+\hat{\mu}))^3\\
\nonumber
&\hspace{11pt}
-\frac{3l}{N+1+l}(z(x)\cdot z^\ast(x+\hat{\mu}))^{l-1}(z^\ast(x) \cdot z(x+\hat{\mu}))^2\\
\nonumber
&\hspace{11pt}
+\frac{3l(l-1)}{(N+1+l)(N+l)}(z(x)\cdot z^\ast(x+\hat{\mu}))^{l-2}(z^\ast(x) \cdot z(x+\hat{\mu}))\\
&\hspace{11pt}
-\frac{l(l-1)(l-2)}{(N+1+l)(N+l)(N-1+l)}(z(x)\cdot z^\ast(x+\hat{\mu}))^{l-3}\Bigr].
\label{fl3}
\end{align}

%\end{comment}

\section{Some examples of $F$ and $\tilde{F}$}\label{FFt}

\noindent
For $m=0$, 
\begin{align}
F^{a_1,\cdots,a_{l}}_{(l;0)}(z)
=
C_{(l;0)}z^{a_1}\cdots z^{a_l},
\label{Fl0}
\end{align}
\begin{align}
\tilde{F}^{a_1,\cdots,a_{l}}_{(l;0)}(z)
=
C_{(l;0)}z^{\ast a_1}\cdots z^{\ast a_l},
\label{tildeFl0}
\end{align}
with
\begin{align}
C_{(l;0)}\equiv
\left(\frac{(N-1+l)!}{l!(N-2)!(N-1)}\right)^\frac{1}{4}.
\end{align}

\noindent
For $m=1$,
\begin{align}
F^{a_1,\cdots,a_{l},a'_1}_{(l;1)}(z)
=
C_{(l;1)}\Big[z^{a_1} z^{\ast a'_1} +E_{(l;1)}^{a_1 a'_1}\Big]
z^{a_2} \cdots z^{a_l},
\label{Fl1}
\end{align}
\begin{align}
\tilde{F}^{a_1,\cdots,a_{l},a'_1}_{(l;1)}(z)
=
C_{(l;1)}\Big[z^{\ast a_1} z^{a'_1} +\tilde{E}_{(l;1)}^{a_1 a'_1}\Big]
z^{\ast a_2}\cdots z^{\ast a_l},
\label{tildeFl1}
\end{align}
with
\begin{align}
C_{(l;1)}
=
\left(\frac{(N+l)!(N-1+l)}{l!(N-1)!(N-1)}\right)^\frac{1}{4},
\end{align}
\begin{align}
E_{(l;1)}^{a_1 a'_1}
=
\sqrt{\frac{l}{N(N-1+l)}}\delta^{a_1 a'_1},
\end{align}
\begin{align}
\tilde{E}_{(l;1)}^{a_1 a'_1}
=
-\sqrt{\frac{l}{N(N-1+l)}}\delta^{a_1 a'_1}.
\end{align}

\noindent
For $m=2$,
\begin{align}
\nonumber
&F^{a_1,\cdots,a_{l},a'_1,a'_2}_{(l;2)}(z)\\
&=C_{(l;2)}
\Big[z^{a_1} z^{a_2} z^{\ast a'_1} z^{\ast a'_2}
+E^{a_1 a'_1}_{(l;2)}z^{a_2}z^{\ast a'_2}\Big]
z^{a_3}\cdots z^{a_l},
\label{Fl2}
\end{align}

\begin{align}
\nonumber
&\tilde{F}^{a_1,\cdots,a_{l},a'_1,a'_2}_{(l;2)}(z)\\
&= C_{(l;2)}
\Big[ z^{\ast a_1} z^{\ast a_2} z^{a'_1} z^{a'_2}
+\tilde{E}^{a_2 a'_2}_{(l;2)}z^{\ast a_1} z^{a'_1}\Big]z^{\ast a_3}\cdots z^{\ast a_l},
\label{tildeFl2}
\end{align}
with
\begin{align}
C_{(l;2)}=
\left(\frac{(N+1+l)!(N+l)(N-1+l)}{2!l!N!(N-1)}\right)^\frac{1}{4},
\end{align}
\begin{align}
E_{(l;2)}^{a_1 a'_1}
=
-\frac{l(N-1+l)+\sqrt{lN(N-1+l)}}{(N+l)(N-1+l)}\delta^{a_1 a'_1},
\end{align}
\begin{align}
\tilde{E}_{(l;2)}^{a_2 a'_2}
=
-\frac{l(N-1+l)-\sqrt{lN(N-1+l)}}{(N+l)(N-1+l)}\delta^{a_2 a'_2}.
\end{align}

%\begin{comment}

\noindent
For $m=3$,
\begin{align}
\nonumber
&F^{a_1,\cdots,a_{l},a'_1,a'_2,a'_3}_{(l;3)}(z)\\
&=C_{(l;3)}
\Big[z^{a_1} z^{a_2} z^{a_3} z^{\ast a'_1} z^{\ast a'_2} z^{\ast a'_3}
+E^{a_1 a'_1}_{1(l;3)}z^{a_2}z^{a_3}z^{\ast a'_2}z^{\ast a'_3}
+E^{a_1 a_3 a'_1 a'_3}_{2(l;3)}z^{a_2}z^{\ast a'_2}\Big]
z^{a_4}\cdots z^{a_l},
\label{Fl3}
\end{align}

\begin{align}
\nonumber
&\tilde{F}^{a_1,\cdots,a_{l},a'_1,a'_2,a'_3}_{(l;3)}(z)\\
&= C_{(l;3)}
\Big[ z^{\ast a_1} z^{\ast a_2} z^{\ast a_3} z^{a'_1} z^{a'_2} z^{a'_3}
+\tilde{E}^{a_2 a'_2}_{1(l;3)}z^{\ast a_1} z^{\ast a_3} z^{a'_1} z^{a'_3}
+\tilde{E}^{a_2 a_3 a'_2 a'_3}_{2(l;3)}z^{\ast a_1} z^{a'_1}\Big]
z^{\ast a_4}\cdots z^{\ast a_l},
\label{tildeFl3}
\end{align}
with
\begin{align}
C_{(l;3)}=
\left(\frac{(N+2+l)!(N+1+l)(N+l)(N-1+l)}{3!l!(N+1)!(N-1)}\right)^\frac{1}{4},
\end{align}

\begin{align}
E_{1(l;3)}^{a_1 a'_1}
=
\frac{-\frac{3l}{N+1+l}+\sqrt{X}}{2}\delta^{a_1 a'_1},
\end{align}
\begin{align}
\tilde{E}_{1(l;3)}^{a_2 a'_2}
=
\frac{-\frac{3l}{N+1+l}-\sqrt{X}}{2}\delta^{a_2 a'_2},
\end{align}

\begin{align}
E_{2(l;3)}^{a_1 a_3 a'_1 a'_3}
=
\frac{\sqrt{Y}
-\sqrt{X}}{2N}
\delta^{a_1 a'_1}\delta^{a_3 a'_3},
\end{align}
\begin{align}
\tilde{E}_{2(l;3)}^{a_2 a_3 a'_2 a'_3}
=
-\frac{\sqrt{Y}
-\sqrt{X}}{2N}
\delta^{a_2 a'_2}\delta^{a_3 a'_3},
\end{align}

\begin{align}
X=\frac{3l(N+4-Nl-l^2)}{(N+l)(N+1+l)^2},
\end{align}
\begin{align}
Y=X+\frac{4Nl(l-1)(l-2)}{(N+1+l)(N+l)(N-1+l)}.
\end{align}

%\end{comment}

\section{Some examples of the tensor $D$}\label{Dtensor}

In this appendix, we show several examples of the tensor $D^x$ in Eq. (\ref{dtensor}).
Here, we define
\begin{align}
W_{(l;m)}
\equiv
\sqrt{d_{(l;m)}h_{(l;m)}(\beta)}C_{(l;m)}.
\end{align}

\noindent
Example 1:
\begin{align}
\nonumber
&D^x_{((l_s ; 0),\{a\})((l_t ; 0),\{b\})((l_u ; 0),\{c\})((l_v ; 0),\{d\})}\\
\nonumber
&=\int dz(x)
\sqrt{d_{(l_s;0)}d_{(l_t;0)}d_{(l_u;0)}d_{(l_v;0)}
h_{(l_s;0)}(\beta)h_{(l_t;0)}(\beta)
h_{(l_u;0)}(\beta)h_{(l_v;0)}(\beta)}\\
\nonumber
&\hspace{60pt}
\times
\tilde{F}^{a_1,\cdots,a_{l_s}}_{(l_s;0)}(z(x))
F^{b_1,\cdots,b_{l_t}}_{(l_t;0)}(z(x))
\tilde{F}^{c_1,\cdots,c_{l_u}}_{(l_u;0)}(z(x))
F^{d_1,\cdots,d_{l_v}}_{(l_v;0)}(z(x))\\
\nonumber
&=W_{(l_s;0)}W_{(l_t;0)}W_{(l_u;0)}W_{(l_v;0)}\\
\nonumber
&\hspace{11pt}
\times
\int dz(x)
[z^{\ast a_1}(x)\cdots z^{\ast a_{l_s}}(x)][z^{b_1}(x)\cdots z^{b_{l_t}}(x)]
[z^{\ast c_1}(x)\cdots z^{\ast c_{l_u}}(x)][z^{d_1}(x)\cdots z^{d_{l_v}}(x)]\\
&=W_{(l_s;0)}W_{(l_t;0)}W_{(l_u;0)}W_{(l_v;0)}
\delta^{l_s+l_u}_{l_t+l_v}\frac{(N-1)!}{(N-1+l_s+l_u)!}
\delta^{\{a_1,\cdots, a_{l_s},c_1,\cdots, c_{l_u}\}}
_{\{b_1,\cdots , b_{l_t},d_1,\cdots ,d_{l_v}\}}.
\end{align}
In the last two lines, Eqs. (\ref{Fl0}) and (\ref{zint}) are used.

\noindent
Example 2:
\begin{align}
\nonumber
&D^x_{((l_s;1),\{a\})((l_t;0),\{b\})((l_u;0),\{c\})((l_v;0),\{d\})}\\
\nonumber
&=W_{(l_s;1)}W_{(l_t;0)}W_{(l_u;0)}W_{(l_v;0)}\\
\nonumber
&\hspace{11pt}\times
\int dz(x)
\Big[z^{\ast a_1}(x) z^{a'_1}(x) +\tilde{E}_{(l_s;1)}^{a_1 a'_1}\Big]
[z^{\ast a_2}(x)\cdots z^{\ast a_{l_s}}(x)]
[z^{b_1}(x)\cdots z^{b_{l_t}}(x)]\\
\nonumber
&\hspace{70pt}
\times
[z^{\ast c_1}(x)\cdots z^{\ast c_{l_u}}(x)][z^{d_1}(x)\cdots z^{d_{l_v}}(x)]\\
\nonumber
&=W_{(l_s;1)}W_{(l_t;0)}W_{(l_u;0)}W_{(l_v;0)}\\
\nonumber
&\hspace{10pt}\times
\Big[
\delta^{l_s+l_u}_{l_t+l_v+1}\frac{(N-1)!}{(N-1+l_s+l_u)!}
\delta^{\{a_1,\cdots , a_{l_s},c_1,\cdots ,c_{l_u}\}}
_{\{a'_1,b_1,\cdots , b_{l_t},d_1,\cdots ,d_{l_v}\}}\\
&\hspace{20pt}
+
\tilde{E}_{(l_s;1)}^{a_1 a'_1}
\delta^{l_s+l_u-1}_{l_t+l_v}\frac{(N-1)!}{(N-1+l_s+l_u-1)!}
\delta^{\{a_2,\cdots , a_{l_s},c_1,\cdots , c_{l_u}\}}
_{\{b_1,\cdots , b_{l_t},d_1,\cdots ,d_{l_v}\}}
\Big].
\end{align}

\noindent
Example 3:
\begin{align}
\nonumber
&D^x_{((l_s;2),\{a\})((l_t;0),\{b\})((l_u;0),\{c\})((l_v;0),\{d\})}\\
\nonumber
&=W_{(l_s;2)}W_{(l_t;0)}W_{(l_u;0)}W_{(l_v;0)}\\
\nonumber
&\hspace{11pt}\times
\int dz(x)
\Big[ z^{\ast a_1}(x) z^{\ast a_2}(x) z^{a'_1}(x) z^{a'_2}(x)
+\tilde{E}^{a_2 a'_2}_{(l_s;2)}z^{\ast a_1}(x)z^{a'_1}(x)\Big]
[z^{\ast a_3}(x)\cdots z^{\ast a_{l_s}}(x)]\\
\nonumber
&\hspace{70pt}
\times
[z^{b_1}(x)\cdots z^{b_{l_t}}(x)]
[z^{\ast c_1}(x)\cdots z^{\ast c_{l_u}}(x)][z^{d_1}(x)\cdots z^{d_{l_v}}(x)]\\
\nonumber
&=W_{(l_s;2)}W_{(l_t;0)}W_{(l_u;0)}W_{(l_v;0)}\\
\nonumber
&\hspace{10pt}\times
\Big[
\delta^{l_s+l_u}_{l_t+l_v+2}\frac{(N-1)!}{(N-1+l_s+l_u)!}
\delta^{\{a_1,\cdots , a_{l_s},c_1,\cdots , c_{l_u}\}}
_{\{a'_1,a'_2,b_1,\cdots , b_{l_t},d_1,\cdots ,d_{l_v}\}}\\
&\hspace{20pt}
+
\tilde{E}_{(l_s;2)}^{a_2 a'_2}
\delta^{l_s+l_u-1}_{l_t+l_v+1}\frac{(N-1)!}{(N-1+l_s+l_u-1)!}
\delta^{\{a_1,a_3,a_4,\cdots , a_{l_s},c_1,\cdots , c_{l_u}\}}
_{\{a'_1,b_1,\cdots , b_{l_t},d_1,\cdots ,d_{l_v}\}}
\Big].
\end{align}

\begin{align}
\nonumber
&D^x_{((l_s;0),\{a\})((l_t;2),\{b\})((l_u;0),\{c\})((l_v;0),\{d\})}\\
\nonumber
&=W_{(l_s;0)}W_{(l_t;2)}W_{(l_u;0)}W_{(l_v;0)}\\
\nonumber
&\hspace{11pt}\times
\int dz(x)
[z^{a_1}(x)\cdots z^{\ast a_{l_s}}(x)]
\Big[ z^{b_1}(x) z^{b_2}(x) z^{\ast b'_1}(x) z^{\ast b'_2}(x)
+E^{b_1 b'_1}_{(l_t;2)}z^{b_2}(x)z^{\ast b'_2}(x)\Big]\\
\nonumber
&\hspace{70pt}
\times
[z^{b_3}(x)\cdots z^{b_{l_t}}(x)]
[z^{\ast c_1}(x)\cdots z^{\ast c_{l_u}}(x)][z^{d_1}(x)\cdots z^{d_{l_v}}(x)]\\
\nonumber
&=W_{(l_s;0)}W_{(l_t;2)}W_{(l_u;0)}W_{(l_v;0)}\\
\nonumber
&\hspace{10pt}\times
\Big[
\delta^{l_s+l_u+2}_{l_t+l_v}\frac{(N-1)!}{(N-1+l_s+l_u+2)!}
\delta^{\{a_1,\cdots , a_{l_s},b'_1,b'_2,c_1,\cdots , c_{l_u}\}}
_{\{b_1,b_2,\cdots , b_{l_t},d_1,\cdots ,d_{l_v}\}}\\
&\hspace{20pt}
+
E_{(l_t;2)}^{b_1 b'_1}
\delta^{l_s+l_u+1}_{l_t+l_v-1}\frac{(N-1)!}{(N-1+l_s+l_u+1)!}
\delta^{\{a_1,\cdots , a_{l_s},b'_2,c_1,\cdots , c_{l_u}\}}
_{\{b_2,\cdots , b_{l_t},d_1,\cdots ,d_{l_v}\}}
\Big].
\end{align}

\noindent
For more simplicity, we define
\begin{align}
\nonumber
&S^x_{\widetilde{((l_s;m_s),\{a\})}((l_t;m_t),\{b\})\widetilde{((l_u;m_u),\{c\})}((l_v;m_v),\{d\})}\\
&\hspace{30pt}
\equiv
\delta^{l_s+m_t+l_u+m_v}_{m_s+l_t+m_u+l_v}\frac{(N-1)!}{(N-1+l_s+m_t+l_u+m_v)!}
\delta^{\{a_1,\cdots , a_{l_s}, b'_1,\cdots , b'_{m_t}, c_1,\cdots , c_{l_u}, d'_1,\cdots ,d'_{m_v}\}}
_{\{a'_1,\cdots , a'_{m_s}, b_1,\cdots , b_{l_t}, c'_1,\cdots , c'_{m_u}, d_1,\cdots ,d_{l_v}\}}.
\end{align}
Using this definition, for example, we can express
\begin{align}
S^x_{\widetilde{((l_s;2),\{a\})}((l_t;0),\{b\})\widetilde{((l_u;0),\{c\})}((l_v;0),\{d\})}
=
\delta^{l_s+l_u}_{l_t+l_v+2}\frac{(N-1)!}{(N-1+l_s+l_u)!}
\delta^{\{a_1,\cdots , a_{l_s},c_1,\cdots , c_{l_u}\}}
_{\{a'_1,a'_2,b_1,\cdots , b_{l_t},d_1,\cdots ,d_{l_v}\}},
\end{align}
\begin{align}
S^x_{\widetilde{((l_s;0),\{a\})}((l_t;2),\{b\})\widetilde{((l_u;0),\{c\})}((l_v;0),\{d\})}
=
\delta^{l_s+l_u+2}_{l_t+l_v}\frac{(N-1)!}{(N-1+l_s+l_u+2)!}
\delta^{\{a_1,\cdots , a_{l_s},b'_1,b'_2,c_1,\cdots , c_{l_u}\}}
_{\{b_1,b_2,\cdots , b_{l_t},d_1,\cdots ,d_{l_v}\}},
\end{align}
\begin{align}
S^x_{\widetilde{((l_s-1;1),\{a\})}((l_t;0),\{b\})\widetilde{((l_u;0),\{c\})}((l_v;0),\{d\})}
=
\delta^{l_s+l_u-1}_{l_t+l_v+1}\frac{(N-1)!}{(N-1+l_s+l_u-1)!}
\delta^{\{a_1,a_3,a_4,\cdots , a_{l_s},c_1,\cdots , c_{l_u}\}}
_{\{a'_1,b_1,\cdots , b_{l_t},d_1,\cdots ,d_{l_v}\}},
\end{align}
\begin{align}
S^x_{\widetilde{((l_s;0),\{a\})}((l_t-1;1),\{b\})\widetilde{((l_u;0),\{c\})}((l_v;0),\{d\})}
=
\delta^{l_s+l_u+1}_{l_t+l_v-1}\frac{(N-1)!}{(N-1+l_s+l_u+1)!}
\delta^{\{a_1,\cdots , a_{l_s},b'_2,c_1,\cdots , c_{l_u}\}}
_{\{b_2,\cdots , b_{l_t},d_1,\cdots ,d_{l_v}\}}.
\end{align}
The Kronecker deltas suggest a possibility of block diagonalization of the tensor $D$. 
Using these tensors $S$, we obtain the following expressions:

\begin{align}
\nonumber
&D^x_{((l_s;2),\{a\})((l_t;0),\{b\})((l_u;0),\{c\})((l_v;0),\{d\})}\\
\nonumber
&=W_{(l_s;2)}W_{(l_t;0)}W_{(l_u;0)}W_{(l_v;0)}\\
&\hspace{10pt}
\times
\Big[
S^x_{\widetilde{((l_s;2),\{a\})}((l_t;0),\{b\})\widetilde{((l_u;0),\{c\})}((l_v;0),\{d\})}
+
\tilde{E}_{(l_s;2)}^{a_2 a'_2}
S^x_{\widetilde{((l_s-1;1),\{a\}_2)}((l_t;0),\{b\})\widetilde{((l_u;0),\{c\})}((l_v;0),\{d\})}\Big],
\end{align}
\begin{align}
\nonumber
&D^x_{((l_s;0),\{a\})((l_t;2),\{b\})((l_u;0),\{c\})((l_v;0),\{d\})}\\
\nonumber
&=W_{(l_s;0)}W_{(l_t;2)}W_{(l_u;0)}W_{(l_v;0)}\\
&\hspace{10pt}
\times
\Big[
S^x_{\widetilde{((l_s;0),\{a\})}((l_t;2),\{b\})\widetilde{((l_u;0),\{c\})}((l_v;0),\{d\})}
+
E_{(l_t;2)}^{b_1 b'_1}
S^x_{\widetilde{((l_s;0),\{a\})}((l_t-1;1),\{b\}_1)\widetilde{((l_u;0),\{c\})}((l_v;0),\{d\})}\Big],
\end{align}
where the subscript ``$2$'' of $\{a\}_2$ means the absence of $a_2$ and $a'_2$ in $\{a\}$ and the same holds for $\{b\}$, $\{c\}$, and $\{d\}$.
This simple expression of the tensor $D$ allows us to compute the more complicated tensor easily.

\noindent
For example,
\begin{align}
\nonumber
&D^x_{((l_s;1),\{a\})((l_t;2),\{b\})((l_u;2),\{c\})((l_v;0),\{d\})}\\
\nonumber
&=W_{(l_s;1)}W_{(l_t;2)}W_{(l_u;2)}W_{(l_v;0)}\\
\nonumber
&\hspace{10pt}\times
\Big[
S^x_{\widetilde{((l_s;1),\{a\})}((l_t;2),\{b\})\widetilde{((l_u;2),\{c\})}((l_v;0),\{d\})}
+
\tilde{E}_{(l_s;1)}^{a_1 a'_1}
S^x_{\widetilde{((l_s-1;0),\{a\}_1)}((l_t;2),\{b\})\widetilde{((l_u;2),\{c\})}((l_v;0),\{d\})}\\
\nonumber
&\hspace{10pt}+
E_{(l_t;2)}^{b_1 b'_1}
S^x_{\widetilde{((l_s;1),\{a\})}((l_t-1;1),\{b\}_1)\widetilde{((l_u;2),\{c\})}((l_v;0),\{d\})}
+
\tilde{E}_{(l_u;2)}^{c_2 c'_2}
S^x_{\widetilde{((l_s;1),\{a\})}((l_t;2),\{b\})\widetilde{((l_u-1;1),\{c\}_2)}((l_v;0),\{d\})}\\
\nonumber
&\hspace{10pt}+
E_{(l_t;2)}^{b_1 b'_1}\tilde{E}_{(l_u;2)}^{c_2 c'_2}
S^x_{\widetilde{((l_s;1),\{a\})}((l_t-1;1),\{b\}_1)\widetilde{((l_u-1;1),\{c\}_2)}((l_v;0),\{d\})}\\
\nonumber
&\hspace{10pt}+
\tilde{E}_{(l_s;1)}^{a_1 a'_1}\tilde{E}_{(l_u;2)}^{c_2 c'_2}
S^x_{\widetilde{((l_s-1;0),\{a\}_1)}((l_t;2),\{b\})\widetilde{((l_u-1;1),\{c\}_2)}((l_v;0),\{d\})}\\
\nonumber
&\hspace{10pt}+
\tilde{E}_{(l_s;1)}^{a_1 a'_1}E_{(l_t;2)}^{b_1 b'_1}
S^x_{\widetilde{((l_s-1;0),\{a\}_1)}((l_t-1;1),\{b\}_1)\widetilde{((l_u;2),\{c\})}((l_v;0),\{d\})}\\
&\hspace{10pt}+\tilde{E}_{(l_s;2)}^{a_1 a'_1}E_{(l_t;2)}^{b_1 b'_1}\tilde{E}_{(l_u;2)}^{c_2 c'_2}
S^x_{\widetilde{((l_s-1;0),\{a\}_1)}((l_t-1;1),\{b\}_1)\widetilde{((l_u-1;1),\{c\}_2)}((l_v;0),\{d\})}
\Big].
\end{align}

\end{appendix}

\end{document}